\documentclass[12pt,a4paper]{article}
\usepackage[british]{babel}
\usepackage[dvips]{epsfig}
\usepackage{amssymb}
\epsfclipon
\title{
  {\vspace{-2cm} \normalsize
     \epsfig{figure=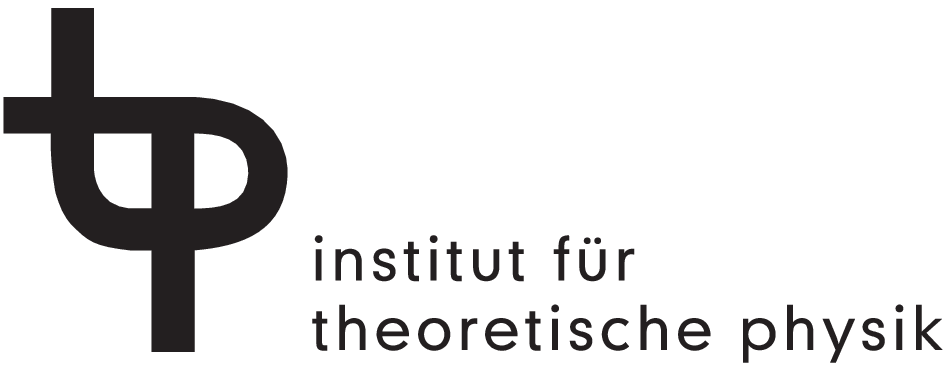,width=80mm}
     \hfill\parbox[b][30mm][t]{35mm}{MS-TP-04-14 \\
                                      hep-lat/0407036}  }\\[15mm]
The low-lying mass spectrum of the N=1 SU(2) SUSY Yang-Mills theory with Wilson fermions}
\author{F.~Farchioni and R.~Peetz\\
        Institut f\"ur Theoretische Physik,
        Universit\"at M\"unster\\
        Wilhelm-Klemm-Str.~9, D-48149 M\"unster, Germany\\
        e-mail: farchion@uni-muenster.de}
\date{July 28, 2004}
\newcommand{\be}{\begin{equation}}
\newcommand{\ee}{\end{equation}}
\newcommand{\bea}{\begin{eqnarray}}
\newcommand{\eea}{\end{eqnarray}}

\newcommand{\Tr}{\mbox{tr}}

\begin{document}
\maketitle

\vspace{-.5cm}

\begin{abstract}
We analyze the low energy spectrum of bound states of the N=1 SU(2)
SUSY Yang-Mills Theory (SYM). This work continues the
investigation of the non-per\-tur\-ba\-tive properties of SYM by Monte
Carlo simulations in the Wilson discretization with dynamical
gluinos. The dynamics of the gluinos is included by the Two-Step
Multi-Bosonic Algorithm (TSMB) for dynamical fermions. A new set
of configurations has been generated on a $16^3\cdot 32$ lattice at
$\beta=2.3$ and $\kappa=0.194$. The analysis also includes sets of
configurations previously generated on a smaller ($12^3\cdot 24$)
lattice at $\kappa=0.1925, 0.194$ and 0.1955. Guided by predictions from
low energy Lagrangians, we consider spin-1/2, scalar and
pseudoscalar particles.
The spectrum of SYM is a challenging subject of investigation
because of the extremely noisy correlators. In particular,
meson-like correlators contain disconnected contributions. The
larger time-extention of the $16^3\cdot 32$ lattice allows to observe two-state
signals in the effective mass. Finite-volume effects are monitored
by comparing results from the two lattice sizes.
\end{abstract}
\newpage
\section{Introduction}
\label{sec1}

The N=1 SU(N$_c$) SUSY Yang-Mills (SYM) theory is the simplest
instance of a SUSY gauge theory and presently the only one
viable for large scale numerical investigations. It describes N$_c^2$-1 gluons
accompanied by an equal number of fermionic partners (gluinos) in
the same (adjoint) representation of the color group. Veneziano and
Yankielowicz~\cite{VeYa} have shown how the assumption of confinement 
in combination with SUSY strongly constrains the low energy structure of the
theory. The expected degrees of freedom dominating the low energy regime 
are composite operators of the gluon and gluino field which can be arranged
into a chiral superfield. These are: the gluino scalar and
pseudoscalar bilinears $\bar{\lambda}\lambda$,
$\bar{\lambda}\gamma_5\lambda$, the corresponding gluonic
quantities  $F^2$, $\tilde{F}F$, and the spin-1/2 gluino-glue
operator $\Tr_c[F\sigma\lambda]$. However the program of including the
purely gluonic operators (``glueballs'') as dynamical degrees of freedom turns out 
to be non trivial~\cite{FaGaSch,BeMi,CeKnaLou,MeSa}. In~\cite{FaGaSch,CeKnaLou,MeSa} 
the Veneziano-Yankielowicz low energy Lagrangian was extended 
so as to include all the desired low energy states,
which are arranged into two Wess-Zumino supermultiplets.
The authors of~\cite{BeMi} pointed out on the other hand, that fulfillment 
of the program requires dynamical SUSY breaking and
its consequent absence from the particle spectrum.
In a situation where the theoretical framework seems to be still unsettled,
a first-principles approach is welcome. This can be provided 
by lattice computations. 

Our goal is to verify the low energy spectrum of SYM in the case of
SU(2) gauge group by numerical techniques. By doing this we
continue past projects, see~\cite{Montvay_SYM} for a review. 
The direct approach to the spectrum of SYM consists in studying the
time-dependence of correlators of operators having the expected quantum numbers
of the low-lying particles. The simplest operators of this type are
the glueball, gluino-glue and mesonic operators also entering the
low energy Lagrangians.
Since gluino bilinears and glueball operators of the same parity carry 
the same (conserved) quantum numbers of the theory, it is natural to expect
mixing among them~\cite{FaGaSch}. We have to stress here that
when the dynamics of the gluinos is taken into account beyond
the valence picture, the disentanglement of the
``unmixed'' states with identical quantum numbers is not possible: 
only the mixed physical states can be the object
of investigation.\footnote{In order to avoid confusion with the mass
pattern of QCD we refrain to associate any name to the particle
states of SYM and will refer to them according to their quantum
numbers (spin and parity).} The result is the determination
of the mass of the lightest particle with the same quantum numbers of
the projecting operator: from this point of view glueball and mesonic 
operators are equivalent.

The action is discretized in the Wilson 
fashion~\cite{CuVe}\footnote{First simulations with domain wall fermions
were performed in~\cite{Ohio}.} where, however, the gluino is a
Majorana spinor in the adjoint representation: \be
S=S_G[U]+S_f[U,\bar{\lambda},\lambda]\ ; \ee $S_G[U]$ is the usual
plaquette action and \bea S_f[U,\bar{\lambda},\lambda] =
\frac{1}{2} \sum_x \bar{\lambda}(x)\lambda(x)- \frac{\kappa}{2}
\sum_x \sum_\mu
 [\bar{\lambda}(x+\hat{\mu})V_\mu(x)(r+\gamma_\mu)\lambda(x) \nonumber \\ +
 \bar{\lambda}(x)V_\mu^T(x)(r-\gamma_\mu)\lambda(x+\hat{\mu})] \ ;
\label{eq:defLatticeFermionAction} \eea $r$ is the Wilson
parameter set to $r=1$ in our case. The gluino field satisfies the
Majorana condition 
\be 
\lambda=\lambda^{\cal C}={\cal C} \bar{\lambda}^T\ ,
\ee 
where the charge conjugation in the spinorial
representation is ${\cal C}=\gamma_0\gamma_2$; the gauge link in
the adjoint representation reads: \be [V_\mu(x)]_{ab} \equiv 2 \Tr
[U_\mu^\dag(x)T^aU_\mu(x)T^b] = [V_\mu^*(x)]_{ab} =
[V_\mu^T(x)]^{-1}_{ab}\ , \label{defAdjointRep} \ee where $T^a$
are the generators of the color group.

The dynamics of the gluinos is included by adopting the Two-Step
Multi-Bosonic Algorithm (TSMB) for dynamical fermions~\cite{Montvay_TSMB}. 
The algorithm has the nice feature of
accommodating any, even fractional, number of flavors. This is
required for SYM since, schematically, the gluino has only half of
the degrees of freedom of a Dirac fermion and consequently the
fermion measure contains the square root of the fermion
determinant: this corresponds to $N_f=1/2$.
In addition (cf.~\cite{CaetAl} for details) the design of TSMB is
optimized to deal with light fermionic degrees of freedom,
a critical factor when approaching the SUSY limit.
Tests of the algorithm performance in QCD for light quark masses
can be found in~\cite{QCD_TSMB}.

In the Wilson discretization SUSY is broken in a
two-fold way: explicitly by the Wilson term ensuring the correct
balance between fermionic and bosonic degrees of freedom in the
continuum limit, and softly by the gluino mass term. On the basis
of the Ward identities~\cite{CuVe, DoetAl, FaetAl}, SUSY is
expected to be recovered in the continuum limit by tuning the
gluino mass to zero. (The situation is perfectly analogous to that
of QCD, where chirality is recovered by tuning the quark mass to
zero). However, $O(a)$ and $O(m_{\tilde{g}})$ SUSY violating
effects are expected to distort the SUSY pattern in
practical situations. A systematic analytical expansion in the gluino mass
is missing in SYM, therefore it is not obvious how to set the scale for the $O(m_{\tilde{g}})$
breaking (something analogous to $\Lambda_\chi=4\pi f_\pi$ in
chiral perturbation theory). 
The only possibility, at least for the moment, to gain some information on 
the effective ``heaviness'' of the gluino is to force analogy with QCD. 
Needless to say, this procedure is only of
heuristic value. The strategy we adopt is to gradually
increase the hopping parameter $\kappa$ in the Wilson action at fixed value
of the gauge coupling $\beta=2.3$ corresponding to a fairly small
lattice size in QCD units ($a\approx 0.06$ fm), pushing the
simulation towards a lighter and lighter gluino.

First large scale simulations of SYM were performed in~\cite{CaetAl}
on a $12^3\cdot  24$ for $\kappa =0.1925$. New sets of
configurations were produced in~\cite{FaetAl} for $\kappa=0.194$, 0.1955.
We now turn to a $16^3\cdot  32$ lattice, whose larger time extention
allows for a better analysis of the spectrum. We consider here $\kappa=0.194$
(simulations at $\kappa=0.1955, 0.196$ are in progress).
The larger space extention allows us to monitor finite-volume effects in the spectrum.

The spectrum of SYM is challenging from the point of view of
numerical analysis. The signal for the correlators of purely gluonic operators vanishes very rapidly
(ideally one should use anisotropic lattices). The mixed
gluonic-fermionic operators, typical for SUSY models, receive
substantial fluctuations from the gluonic content. A
better asymptotic behavior of the effective mass, however, can be obtained by
combined smearing of the fermionic and gluonic degrees of freedom.
Finally for mesonic operators, special techniques are required for the disconnected term
in the correlator. Here we employ stochastic estimators (SET)~\cite{SET}. Also, we
apply an improved version~\cite{FaMuPe} of the Volume Source
Technique (VST)~\cite{VST}. For fermions in the real
representation of the gauge group, as is the case for SYM, the
original formulation in~\cite{VST} cannot be used. The two
independent techniques were tested in a comparative study for SYM
in~\cite{FaMuPe}.

Description and some results of this study have been reported in~\cite{RoPe}.

The plan of the paper is as follows. In Section 2 we report details of the
simulations and characterize the gauge sample, using analogy with QCD,
by the Sommer scale parameter $r_0$ and the pseudo-pion mass; the gluino mass
is obtained from the soft-breaking term in the SUSY Ward identities;
Section 3 contains methodology and results for the spectrum; in Section 4 we discuss
results and indicate possible directions of improvement.

\begin{table}[t]
  \centering
  \begin{tabular}{cllrrcll}\hline
    reference & $L_s$  & \multicolumn{1}{c}{$\kappa$} &
    \multicolumn{1}{c}{$N_{config}$} & \multicolumn{1}{c}{$N_{cycle}$}
    & $N_{lat}$ & \multicolumn{1}{c}{$r_0/a$} & $a\sqrt{\sigma}$
    \\
\hline
\cite{CaetAl} &  $8$  & 0.19 & 20768 & 1038400 & 32 & 5.41(28)\ \cite{CaetAl} & 0.22(1)\\
\cite{CaetAl} &  $12$ &0.1925& 4320 & 216000 &    9  & 6.71(19)\ \cite{FaetAl} & 0.176(4)\\
\cite{FaetAl} &  $12$ &0.194& 2034 & 42030&    9  & 7.37(30)\ \cite{FaetAl} & 0.160(6)\\
 this study  & $16$ &0.194& 3890 & 25650 &    4  & 7.16(25) & 0.165(9)\\ 
 \cite{FaetAl} &  $12$ &0.1955& 4272 &65832 &    8  & 7.98(48)\ \cite{FaetAl} & 0.147(8)\\
 \hline
\end{tabular}
\caption{Overview of the ensembles used in this work with determination of 
  Sommer scale parameter and string tension.
  The fourth and fifth column respectively report the total number of
  configurations and of update cycles at equilibrium, the sixth column the
  number of replica lattices.
  \label{tab:ensembles}}
\end{table}


\begin{figure}[ht]
  \centerline{\epsfig{file=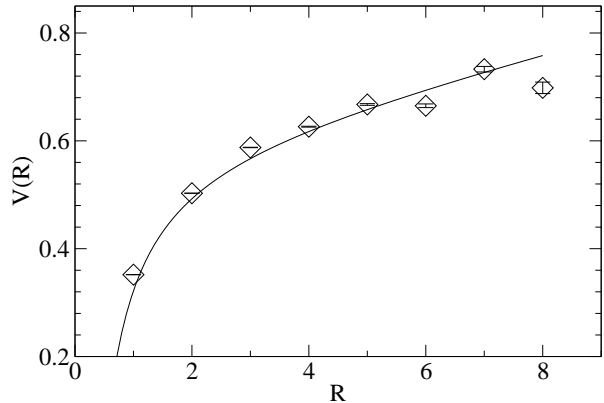, angle=-90,width=11cm}}
  \vspace{-1.5cm}
  \caption{Static potential between heavy sources in the fundamental
  representation on the $16^3\cdot  32$ lattice, $\kappa=0.194$; the line is
  the fit with lattice formulae.
  \label{fig:static_pot}}
\end{figure}

\section{The gauge sample}
\label{sec:sample}

The gauge configurations were generated by the Two-Step
Multi-Bosonic Algorithm for dynamical gluinos~\cite{Montvay_TSMB}.
We refer to~\cite{CaetAl} for a description of the algorithm.
Table~\ref{tab:ensembles} reports an overview of the $\beta=2.3$
ensembles used in this work; the set on the $16^3\cdot 32$ lattice
with $\kappa=0.194$ was newly generated. The setup of the TSMB was
as follows. The local part of the updating procedure (one cycle)
consisted of two steps of heat bath for the bosonic fields
followed by two steps of over-relaxation; the updating for the
gauge sector was obtained by 36 Metropolis sweeps. At the end of
each cycle an accept-reject test was performed on the gauge
configuration, along the lines of the general procedure described
in~\cite{CaetAl}. Every five cycles a global heat-bath step was
applied on the bosonic fields. The typical condition number of the
squared hermitian fermion matrix was $\sim 10^4$. The integrated
autocorrelation time for the smallest eigenvalue was $\sim 240$
cycles.

\subsection{Static potential, string tension  and Sommer scale parameter}

We measured the potential between heavy sources in the fundamental
representation. The results on the larger $16^3\cdot  32$ lattice
confirm the picture of confinement found in~\cite{CaetAl}, see
Fig.~\ref{fig:static_pot}. The Sommer scale parameter $r_0$ and
the string tension $\sqrt{\sigma}$ were measured by fitting the
potential with the lattice formula~\cite{Edw98}
\be
V({\bf r})=V_0+\sigma r-4\pi e \int_{-\pi}^{\pi}\frac{\mbox{d}^3k}{(2\pi)^3}
\frac{\cos({\bf k} \cdot {\bf r})}{4\sum_{j=1}^3\sin^2(k_j/2)}\ ;
\ee
$r_0$ is given by
\be
r_0=\sqrt{\frac{1.65-e}{\sigma}}\ .
\ee
The results are reported in Table~\ref{tab:ensembles}.

Comparing the results on the two lattices, no finite-size effect
beyond statistical uncertainty is visible in the Sommer scale
parameter and the string-tension. Similarly to QCD, the Sommer
scale parameter displays a sizeable gluino-mass dependence. A
linear extrapolation to zero gluino mass is performed in the next
subsection.

\subsection{Massless gluino limit}

\begin{figure}
  \centering
  \centerline {\epsfig{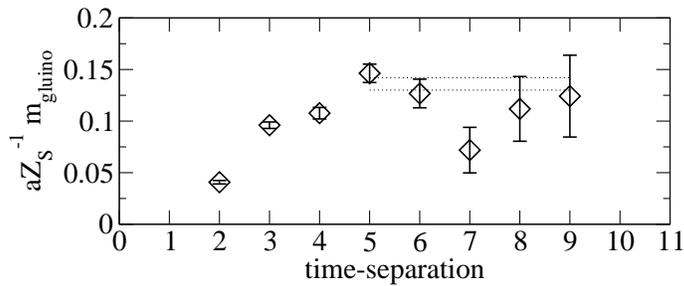}}
  \caption{The gluino mass as obtained from the SUSY Ward identities as a
  function of the time separation between current and insertion operator on the
  $16^3\cdot  32$ lattice, $\kappa=0.194$. The lines indicate bounds of the fit.
  \label{fig:m_gl_vs_ts}}
\end{figure}

In QCD the massless quark limit can be determined
by  inspection of the pion mass or use of the chiral Ward
identities. In contrast, in SYM the $U(1)$ chiral symmetry is
anomalous and the particle with the quantum numbers of the chiral
current (namely the pseudoscalar particle) picks up a mass by the anomaly.
However, theoretical arguments~\cite{VeYa} (cf. also the
discussion in~\cite{DoetAl}) support the picture that the anomaly
is originated by OZI-rule violating diagrams, while the remaining ones
determine spontaneous breaking of the chiral symmetry. The diagrams of the
pseudoscalar correlator respecting the OZI-rule give rise to the
connected (one loop) term, corresponding in QCD to the pion-correlator. The
analogy with QCD suggests the name ``adjoint-pion'' ($a-\pi$) for
the associated pseudoparticle: in the above picture this is
expected to be a soft-mode of the theory, the corresponding mass
disappearing for $m_{\tilde{g}}\rightarrow 0$.

The gluino mass can be directly determined by studying the lattice SUSY
Ward identities~\cite{CuVe,DoetAl,FaetAl}, where the former
enters the soft breaking term. We refer to~\cite{FaetAl} for the
illustration of the method and discussion of theoretical aspects.
One can determine the combination $aZ_S^{-1}m_{\tilde{g}}$
where $Z_S$ is the renormalization constant of the SUSY current,
which is expected to be a (finite) function of the gauge coupling
only. This quantity was determined in~\cite{FaetAl} for the $12^3\cdot 24$
lattice. We repeat here the computation for the $16^3\cdot 32$ lattice.
The results are reported in Fig.~\ref{fig:m_gl_vs_ts}, where the gluino mass is plotted
against the time separation between current and insertion operator
in the SUSY Ward identities. Compared to the $12^3\cdot  24$ case
of~\cite{FaetAl}, the plateau establishes for larger values
of the time separation (five compared to three); unfortunately, at
these time separations the quality of the signal is already quite
deteriorated.
Table~\ref{tab:masses} contains the determinations of
$aZ_S^{-1}m_{\tilde{g}}$ and $am_{a-\pi}$ in present and past
works. Comparison of $12^3\cdot  24$ and $16^3\cdot  32$ results
at $\kappa =0.194$ reveals a sizeable finite volume effect for the
adjoint-pion mass. It should be noted that the sign of the effect
is opposite to the usual one (however this is no
physical mass). The gluino mass comes in larger on the
larger lattice, however within a 1-$\sigma$ effect.

\begin{table}[t]
  \centering
  \begin{tabular}{cllllll}\hline
    $L_s$ & \multicolumn{1}{c}{$\kappa$} & \multicolumn{1}{c}{$am_{a-\pi}$} &
            \multicolumn{1}{c}{$aZ_S^{-1}m_{\tilde{g}}$} &
            \multicolumn{1}{c}{spin-1/2}  &
            \multicolumn{1}{c}{$0^+$(glueb.)} &
            \multicolumn{1}{c}{$0^-$($\bar\lambda\gamma^5\lambda$)}
    \\\hline
     8&    0.19     & 0.71(2) \cite{CaetAl}  &  &   &  & \\
     12&   0.1925   & 0.550(1)  & 0.166(6)\ \cite{FaetAl} & 0.33(4)    &
                      0.53(10) \cite{Kirch} & 0.52(10) \cite{Kirch} \\
     12&   0.194    & 0.470(4)  & 0.124(6)\ \cite{FaetAl} & 0.49(4)    &
                      0.40(11)  & 0.42(1) \\
     16&   0.194    & 0.484(1)  & 0.137(7)                & 0.43(1)    &
                                & 0.52(2) \\
     12&   0.1955   & 0.253(4)  & 0.053(4)\ \cite{FaetAl} & 0.35(4)    &
                      0.36(4)   &0.24(2) \\\hline
  \end{tabular}
\caption{Quantities determined in this and previous studies: the adjoint-pion mass, gluino
            mass from SUSY Ward identities (with local SUSY current, insertion
            operator $\chi^{(sp)}$, cf. Table~5 in~\cite{FaetAl}), spin-1/2,
            $0^+$ and $0^-$ bound state masses.
\label{tab:masses}}
\end{table}

\begin{figure}[t]
  \centerline
       {\epsfig{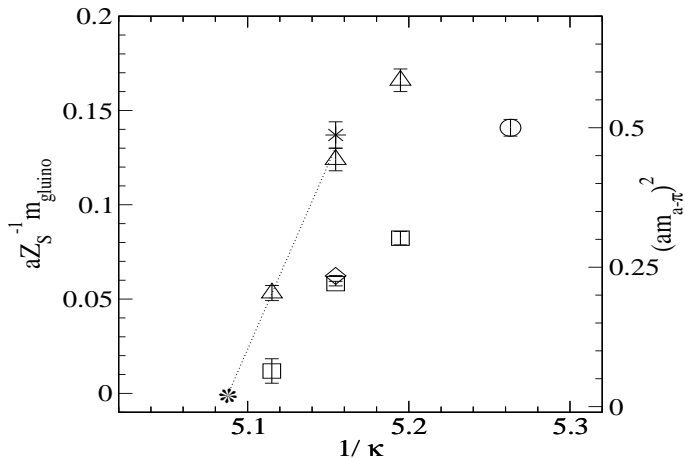}}
\caption{The gluino mass from the SUSY Ward identities and the squared
adjoint-pion mass $m_{a-\pi}$ as a function of $1/\kappa$ (from
the present and past studies~\cite{CaetAl, FaetAl});
$aZ_S^{-1}m_{\tilde g}$ on the $12^3\cdot  24$ lattice
(triangles), the same quantity on the  $16^3\cdot  32$ lattice (star);
 squared adjoint-pion mass on   $8^3\cdot  16$ (circle),
 $12^3\cdot  24$ (boxes), and $16^3\cdot  32$ lattice (diamond). The burst indicates
 the extrapolated massless limit from the two lightest gluino masses.
}
\label{fig:massless_gluino_limit}
\end{figure}

In Fig.~\ref{fig:massless_gluino_limit} $aZ_S^{-1}m_{\tilde{g}}$
is shown together with the squared adjoint-pion mass. 
The two quantities appear to vanish for a common value
of $\kappa\equiv\kappa_c$. The estimate of $\kappa_c$ from the
SUSY Ward identity gluino mass, $\kappa_c\approx 0.1965$~\cite{FaetAl}, is not changed by the inclusion of the point on the
larger lattice.
Using this value of $\kappa_c$ we can now extrapolate the Sommer
scale parameter in Table~\ref{tab:ensembles} to the massless gluino
situation. A linear extrapolation results in 
$r_0/a(m_{\tilde{g}}=0)=8.4(4)$; the error takes into account the
uncertainty in the determination of $\kappa_c$ (assumed to be in
the region  $\kappa$=0.1965-0.1975~\cite{FaetAl}). The Sommer
scale parameter signals the degree of ``smoothness'' (or
``coarseness'') of the gauge sample. In QCD, the present value
would correspond to $a\approx 0.06$ fm (3.3 GeV), a fairly fine
lattice. Further, assuming that the adjoint-pion drives the low
energy features of SYM, as the pion does in QCD, one can estimate
the degree of soft-breaking of SUSY by considering the
dimensionless quantity $M_r=(m_{a-\pi} r_0)^2$. In QCD, validity
of NLO chiral perturbation theory requires~\cite{ShaSho} a
$M_r\lesssim 0.8$ (corresponding to $m_{ud}\lesssim 1/4\, m_s$). In our
case we have $M_r(\kappa=0.194)\approx 16$ and
$M_r(\kappa=0.1955)\approx 4.5$; our lightest case would
correspond in QCD to $m_{ud}\approx 1.5\, m_s$. Alternatively one 
can consider the gluino mass from the  SUSY Ward identity
neglecting $O(1)$ renormalizations, again fixing the scale by the
Sommer parameter with QCD units. In this case we obtain for our
lightest gluino $m_{\tilde{g}}\approx 174$ MeV in rough agreement
with the previous estimate. Since QCD and SU(2) SYM are different
theories, the above indications are of course of qualitative
nature. On the other hand, the relatively large average condition
numbers of the fermion matrix, $\sim 10^4$ for $\kappa=0.194$ and
$\sim 3.6\ 10^4$ for $\kappa=0.1955$, point towards a
lighter gluino.\\

\section{The spectrum}
\label{sec:spectrum}

\begin{figure}[t]
\centerline
       {\epsfig{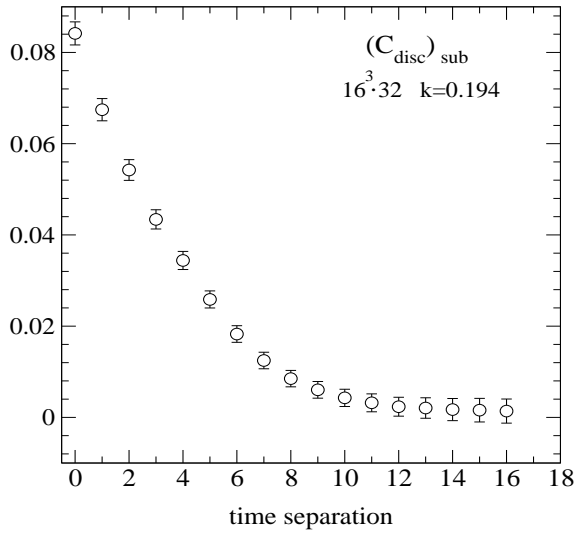}}
\caption{The disconnected component of the  pseudoscalar meson correlator
       after subtraction of the constant term on the $16^3\cdot  32$ lattice, 
$\kappa=0.194$ (SET).}
\label{fig:disc_corr}
\end{figure}

\begin{figure}[t]
\centerline
       {\epsfig{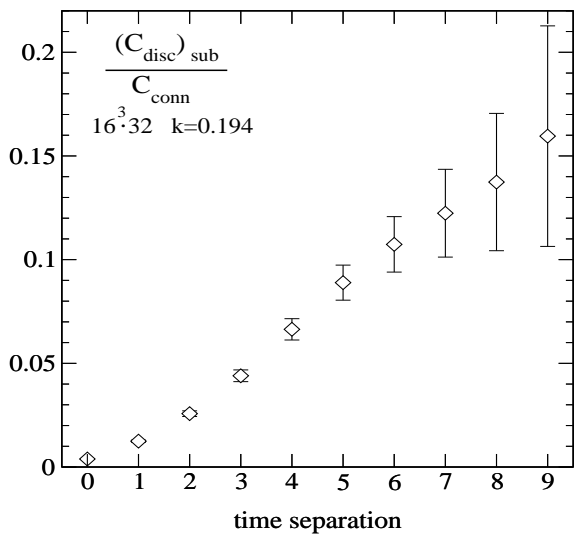}}
\caption{Ratio of the disconnected (after subtraction) and connected components of the  
pseudoscalar meson correlator on the $16^3\cdot  32$ lattice, 
$\kappa=0.194$ (SET).}
\label{fig:ratio_corr}
\end{figure}

As explained above, we concentrate our analysis of the spectrum on
particles with spin=0 (both parities) and spin=1/2. We
investigate the glueball operators, the gluino scalar and pseudoscalar bilinears 
(meson-type operators) and the gluino-glue operator.

\subsection{Spin-1/2 bound states}
\label{sec:gluino-glue}

\begin{figure}
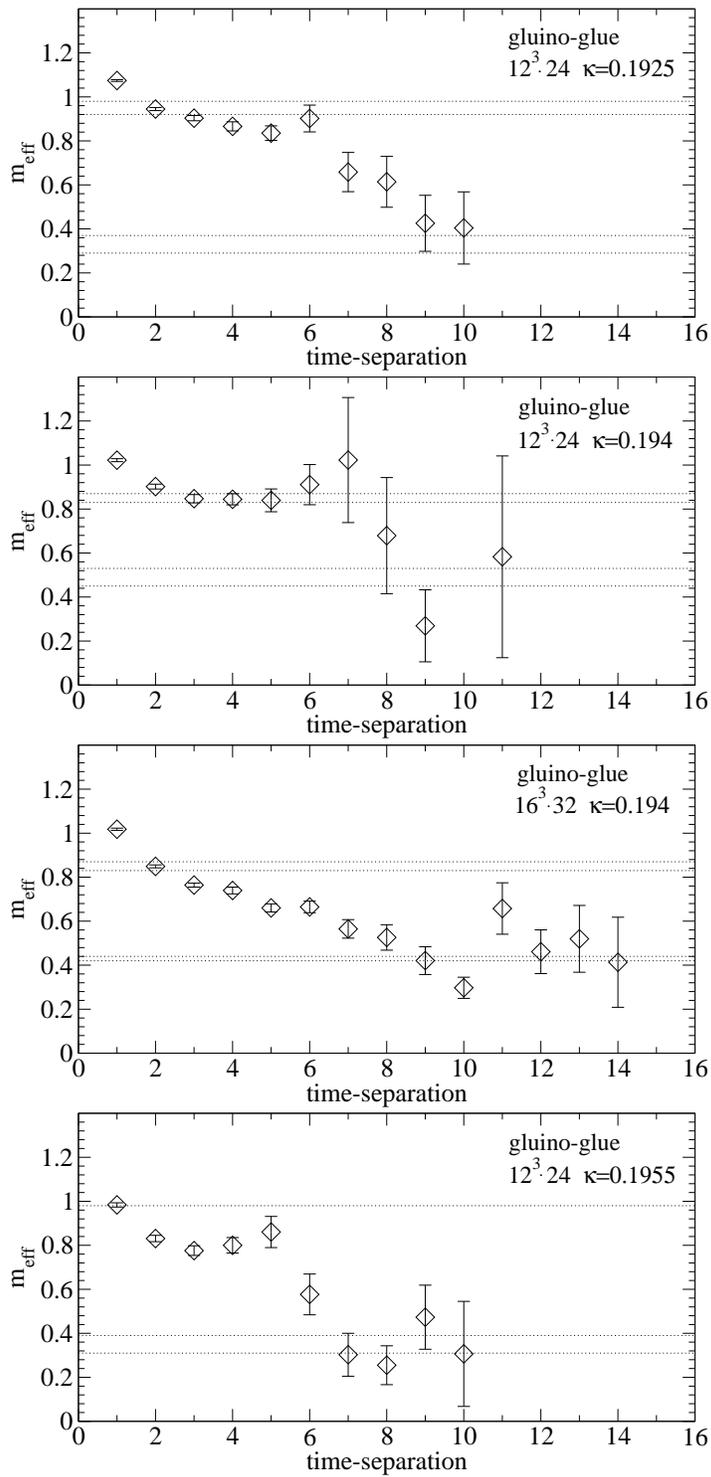

\begin{center}
  \epsfig{file=gg_g4_effmass.1925.12c.eps,width=9.3cm} \\
  \epsfig{file=gg_g4_effmass.194.12c.eps,width=9.3cm}  \\
  \epsfig{file=gg_g4_effmass.194.16c.eps,width=9.3cm}  \\
  \epsfig{file=gg_g4_effmass.1955.12c.eps,width=9.3cm}
\end{center}
\vspace{-.5 cm}
\caption{The effective mass of the spin-1/2 particle ($\gamma_0$ component)
  for the different samples. Dotted lines are the bounds of the mass fits
 (ground and first excited state). In the last case only a rough indication of
 the first excited state mass could be obtained.}
\label{fig:gluino_glue_effmass}
\end{figure}

We adopt here a lattice version~\cite{FaetAl} of the gluino-glue 
operator $\Tr_c[F\sigma\lambda]$ where the
field-strength tensor $F_{\mu\nu}(x)$ is replaced by the
clover-plaquette operator $P_{\mu\nu}(x)$: 
\be 
{\cal O}_{\tilde{g}g}^\alpha(x) = \sum_{i<j}
\sigma_{ij}^{\alpha\beta}\Tr_c[P_{ij}(x)\lambda^\beta(x)]\ ;
\label{eq:defGlGlue} 
\ee 
only spatial indices are taken into
account in order to avoid links in the time-direction. The
clover-plaquette operator is defined to be \be P_{\mu\nu}(x)  =
\frac{1}{8ig_0} \sum_{i=1}^4
\left(U^{(i)}_{\mu\nu}(x)-U^{(i)\dagger}_{\mu\nu}(x)\right) \ee
with \bea U^{(1)}_{\mu\nu} (x) &=& U^\dagger_{\nu}(x)
U^\dagger_{\mu}(x+ \hat{\nu})
 U_{\nu}(x+\hat{\mu}) U_{\mu}(x)\nonumber \\
U^{(2)}_{\mu\nu} (x) &=& U_{\mu}^\dagger(x) U_{\nu}(x-\hat{\nu}+\hat{\mu})
 U_{\mu}(x-\hat{\nu}) U^\dagger_{\nu}(x-\hat{\nu})\nonumber \\
U^{(3)}_{\mu\nu} (x) &=&
 U_{\nu}(x-\hat{\nu}) U_{\nu}(x-\hat{\nu}-\hat{\mu})
 U^\dagger_{\mu}(x-\hat{\nu}-\hat{\mu}) U^\dagger_{\mu}(x-\hat{\mu}) \nonumber \\
U^{(4)}_{\mu\nu} (x) &=& U_{\mu}(x-\hat{\mu}) U^\dagger_{\nu}(x-\hat{\mu})
 U^\dagger_{\mu}(x+\hat{\nu}-\hat{\mu}) U_{\nu}(x)\ .
\label{eq:defClover} \eea The choice of the clover plaquette vs.
the regular plaquette as the gluonic field-strength operator in~(\ref{eq:defGlGlue}) is motivated by the correct behavior under
parity and time reversal transformations as opposed to simply
$U_{\mu\nu}(x)$. Because of the spinorial character of the
gluino-glue, the correlator $C_{\tilde{g}g}(t)$ has a specific
structure in Dirac space. On the basis of the symmetries of the
theory one can show~\cite{DoetAl} that only two components are linearly
independent, $\Tr_D[C_{\tilde{g}g}(x)]$ and $\Tr_D[\gamma_0
C_{\tilde{g}g}(x)]$. In our experience, the latter gives the best
signal. In order to get a better overlap with the ground state, we
apply APE smearing~\cite{Alb87} on the link-variables and Jacobi
smearing~\cite{Jacobi} on the gluino field simultaneously.

\subsection{$0^-$ bound states}
\label{sec:zerominus}

\begin{figure}
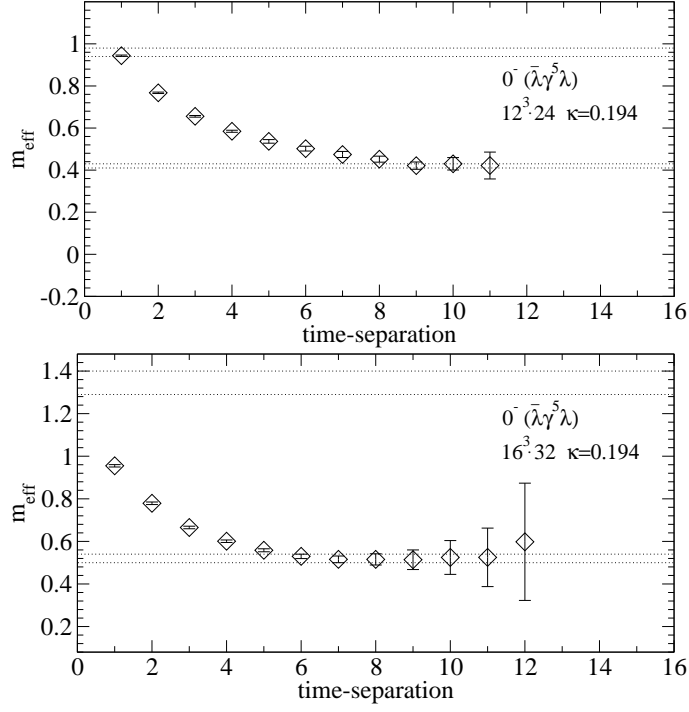

  \begin{center}
    \epsfig{file=a_etap_effmass.194.12c.eps, width=9cm} \\
    \epsfig{file=a_etap_effmass.194.16c.eps, width=9cm}
  \end{center}
  \caption{The comparison of the effective mass of the pseudoscalar particle 
    on the two lattices at $\kappa=0.194$. Dotted lines are the bounds of the mass
    fits (ground and first excited state).}
\label{fig:etap_effmass}
\end{figure}

\begin{figure}
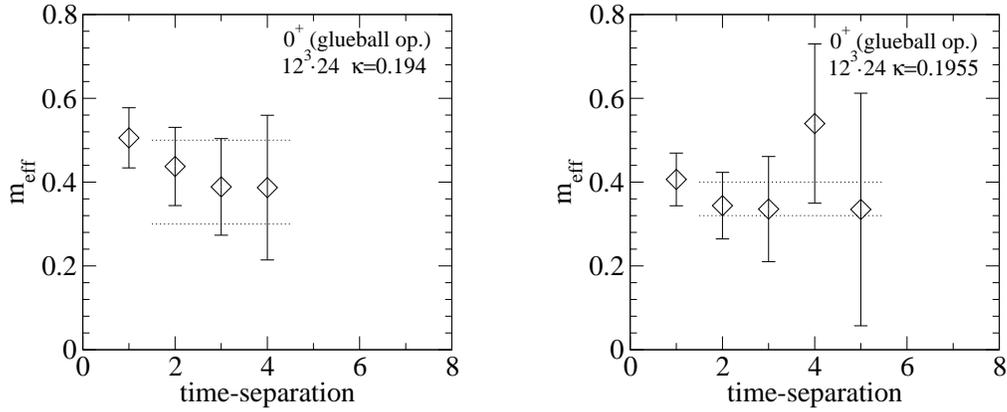

  \begin{center}
    \epsfig{file=glueb_g4_effmass.194.12c.eps, width=6cm} \hspace{1cm}
    \epsfig{file=glueb_g4_effmass.1955.12c.eps, width=6cm}
  \end{center}
\caption{The effective mass of the scalar particle with glueball operator on
    the $12^3\cdot  24$ lattice at $\kappa=0.194, 0.1955$. Dotted lines are
    the bounds of one-mass fits.}
\label{fig:glueb_effmass}
\end{figure}

The meson-type correlators require a separate discussion because
of the disconnected contribution. In the case of SYM one has (with $\Delta$
the gluino propagator): \bea
C_{meson}(x_0-y_0)=C_{conn}(x_0-y_0)-C_{disc}(x_0-y_0)= \hspace{3cm} \nonumber\\
\frac{1}{V_s}\sum_{\vec{x}}
\left<\Tr[\Gamma \Delta_{x,y} \Gamma \Delta_{y,x}]\right>
-\frac{1}{2V_s}\sum_{\vec{x}}\left<\Tr[\Gamma \Delta_{x,x}] \Tr[\Gamma\Delta_{y,y}]
\right>\ ,
  \label{CorrDef}
\eea with $\Gamma \in \{1,\gamma_0\}$ (observe the factor 1/2
reflecting the Majorana nature of the gluino). The disconnected
term requires the estimation of the time-slice sum of the gluino
propagator \be
 S_{\alpha\beta}(x_0) =
\sum_{\vec{x}} \Tr_{c} [\Delta_{x \alpha, x \beta} ]\,.
\label{timeslice} \ee For this, we use the stochastic estimator
technique (SET)~\cite{SET} with complex $Z_2$ noise in the spin
explicit variant SEM~\cite{SEM}. In this case each estimate of the
time-slice sum is obtained by inverting the fermion-matrix with
source $ (\omega_S^{[\alpha]})_{xb\beta} =
\delta_{\alpha\beta}\,\eta^{[\alpha]}_{xb} $ where
$\eta^{[\alpha]}_{xb}$ are independent stochastic variables chosen
at random from $\frac{1}{\sqrt{2}}(\pm 1 \pm i)$. Here we use
point-like operators (i.e.\ no smearing on the gluino). 

On the larger $16^3\cdot 32$ lattice the computed meson correlator displays an offset:
its long-time behavior is not purely exponential, since a constant term also appears. 
Such a constant term is theoretically excluded in the correlator by the symmetries of the
theory. It is present in both SET and VST (see below) determinations of the disconnected
contribution and does not decrease by increasing
the number of the random estimators. In contrast, it is absent in the connected
correlator. We conclude that its origin is to be traced to some cumulative 
numerical effect in the stochastic computation of the disconnected contribution. 

In Fig.~\ref{fig:disc_corr} we show the disconnected component
of the pseudoscalar meson correlator after subtraction of the constant
term. In Fig.~\ref{fig:ratio_corr}
the ratio between the subtracted disconnected component and the connected
one is reported. For a comparison with the same quantity in the
case of QCD see e.g.~\cite{StretAl,CPPACS}; we remark here that
the case of SYM is quite different since the connected correlator
is not related to a physical particle, but rather to the pseudoparticle
$a-\pi$ discussed above.

We cross-check the SET with the improved version~\cite{FaMuPe} of
the Volume Source Technique (VST)~\cite{VST}, applying to fermions
in real representations of the color group. The improvement
consists in averaging the time-slice sums over random gauge
transformations and therefore eliminating the gauge non-invariant
spurious terms. The two methods deliver consistent results of
comparable quality at similar computational cost. For the sake of
brevity, we present here only those from SET.

Another operator with the right quantum numbers ($0^-$) is the
pseudoscalar glueball operator. This is given by a linear
combination of closed loops of link variables which cannot be
rotated into their mirror image (cf. e.g.~\cite{CaetAl}). We
considered the simplest loops of this kind. Unfortunately this
operator does not give a clear signal on our samples.

\subsection{$0^+$ bound states}
\label{sec:zeroplus}

Since the meson-type correlator does not show any appreciable
signal for the $0^+$ state, we turn to the scalar glueball
operator. The standard operator in this case is \be {\cal
O}_{glueball}(x)=\Tr_c[U_{12}(x)+U_{23}(x)+U_{31}(x)]\ . \ee We use
fuzzy operators by applying APE smearing on the link variables.

\subsection{Results}
\label{sec:res}

For all particles we measured the effective masses
(Figs.~\ref{fig:gluino_glue_effmass}-\ref{fig:glueb_effmass}).
In many cases a clear plateau could not be determined. In order to
get a better determination of the ground state mass we used
constrained two-mass fits (bounds in the figures). In some cases
(the spin-1/2 particle on the larger lattice at $\kappa=0.194$,
Fig.~\ref{fig:gluino_glue_effmass},  and the pseudoscalar particle 
at the same $\kappa$ value, Fig.~\ref{fig:etap_effmass})
the two-mass fit can be cross-checked against a plateau of the
effective mass. We ensured the stability of the two-mass fits
by systematically varying fit ranges (for details see~\cite{RoPe}).
The effective mass of the pseudoscalar meson on the $16^3\cdot 32$ lattice,  
lower panel of Fig.~\ref{fig:etap_effmass}, was determined after subtraction
of the constant term in the correlator discussed in Subsection~\ref{sec:zerominus}.

In the case of the scalar glueball
operator, a decrease of the signal/noise ratio was observed on the
larger lattice, as a consequence of which no determination of the
mass was possible.

Results on the determinations of the ground state masses are
reported in Table~\ref{tab:masses} and Fig~\ref{fig:spectrum}. 
A discussion of the results will be presented in the following section.

\section{Discussion}
\label{sec:discussion}

\begin{figure}
  \centerline
{\epsfig{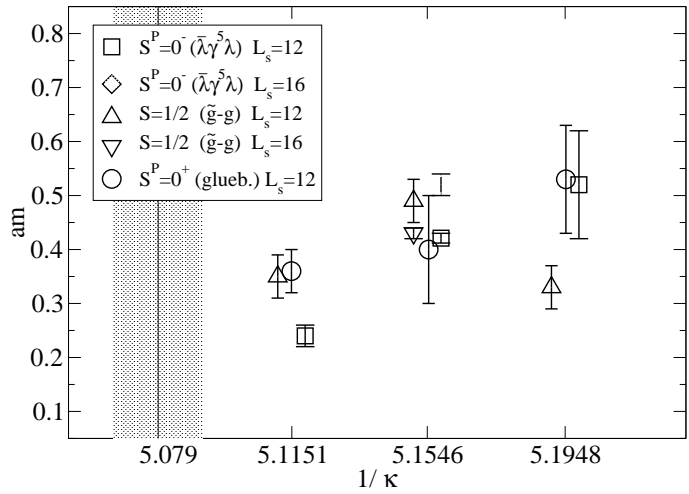}}
\caption{Mass of lightest bound states of SYM determined in this
  work. The shaded region represents the presumed location of the massless gluino
  on the basis of the SUSY Ward identity analysis.}
\label{fig:spectrum}
\end{figure}

Our analysis of the low-lying spectrum of SYM shows a slow
approach of the correlators to the asymptotic behavior where only
the ground state dominates. This is evident in the case of the
gluino-glue and mesonic correlators; in the case of the glueball
correlator, the quality of the signal is not good enough to make
definite statements. Excited states with masses comparable to that
of the ground state are strongly coupled to these operators. In
some cases a plateau of the effective mass emerged and consequently allowed us
to cross-check the results of the two-mass fits. The situation can be improved by
implementing optimized smearing on the operators (in the case of the
gluino bilinears we use point-like operators).

The excited states which hamper the determination of the ground
states are of physical interest by themselves. 
According to~\cite{FaGaSch,CeKnaLou,MeSa}, the first excited states should be
arranged in a second Wess-Zumino supermultiplet. The ``higher
masses'' in our two-mass fits can give a first indication of the
masses of these excited states: the ground state masses lie in
the region 0.2-0.5 (in lattice units), while the higher masses are
in the region 0.8-1. A more refined analysis of the excited states,
however, could be obtained with matrix correlators. In the scalar
sector, one would naturally include the gluino scalar bilinear in
addition to the glueball operator. Given the large fluctuations
observed on the former, the employment of variational methods
would then be advisable. A similar analysis could be done in the
pseudoscalar sector with the corresponding gluino bilinear and the
pseudoscalar glueball operator.

A more fundamental question is whether the employed operators
are optimal in the sense of maximal overlap with the low-lying bound states of SYM.
Investigations could go in the direction of different operators
and different quantum numbers~\cite{JohSchm}.

In the following we restrict the discussion of our results to the
ground states (Fig.~\ref{fig:spectrum}). One of the goals of
this study was to check finite volume effects by comparing
lattices with different spatial extension. This can be done for
our value of $\kappa=0.194$ where two different lattice sizes are
available, $L_s=12$ and $16$. The direct comparison shows, see Table
~\ref{tab:masses}, that a sizeable deviation is present for the pseudoscalar
particle.  Contrary to expectations, the particle comes in heavier on the larger lattice.
For this lattice however an unexpected constant term is observed in the long-time behavior
of the correlator, which could hint at some systematic effect
in the stochastic determination of the disconnected correlator on large
lattices. The pseudoscalar particle is the lightest particle for our
lightest gluino ($\kappa=0.1955$), though in this case only data for the smaller lattice is
available. The scalar and the spin-1/2 particle
have comparable masses, compatible within errors.

Conclusions on the relevance of soft
breaking terms require the control of finite lattice-spacing
effects. Using analogy with QCD in absence of other indications, we
argue that our mesh is relatively fine, while the gluino is still
quite heavy. Next steps will be therefore to consider larger values of
$\kappa$ on large lattices.

\newpage 

\noindent
{\large\bf Acknowledgment}

\noindent

We thank C.~Gebert for participating in the early stages of this work,
I.~Montvay and  G.~M\"unster for stimulating discussions.
The computations were performed on the Cray T3E and JUMP systems at NIC
J\"ulich, the PC clusters at the ZIV of the University of M\"unster
and the Sun Fire SMP-Cluster at the Rechenzentrum of the RWTH Aachen.


\end{document}